\documentclass[final,5p,times,twocolumn]{elsarticle}
 \biboptions{comma,sort&compress}
\usepackage{color}
\usepackage{graphicx}
\usepackage{here}
\usepackage[dvips]{epsfig}
\def\beq{\begin{equation}}
\def\eeq{\end{equation}}
\def\bea{\begin{eqnarray}}
\def\eea{\end{eqnarray}}
\def\bq{\begin{quote}}
\def\eq{\end{quote}}

\def\nnb{\nonumber}
\def\ga{\left(}
\def\dr{\right)}

\def\lrar{\Longrightarrow}

\def\lr{\leftrightarrow}		
\def\nnb{\nonumber}
\def\la{\langle}
\def\ra{\rangle}
\def\nin{\noindent}
\def\ba{\vspace*{-0.2cm}\begin{array}}
\def\ea{\end{array}\vspace*{-0.2cm}}

\def\b{$\bullet~$}
\def\als{\alpha_s}

\def\gg2{ \la\alpha_s G^2 \ra}
\def\gg3{g^3f_{abc}\la G^aG^bG^c \ra}
\def\ggg4{\la\als^2G^4\ra}

\def\beq{\begin{equation}}
\def\enq{\end{equation}}
\def\beqa{\begin{eqnarray}}
\def\enqa{\end{eqnarray}}
\def\nnb{\nonumber}

\journal{Elsevier}

\begin{document}

\begin{frontmatter}

\title{ SVZ $\oplus$ 1/$q^2$ expansion versus some QCD holographic Models}
 \author[label1]{F. Jugeau}
\ead{frederic.jugeau@if.ufrj.br}
\address[label1]{Instituto de F\'{\i}sica, Universidade Federal do Rio de Janeiro, Caixa Postal 68528, RJ 21941-972, Rio de Janeiro, Brazil.}
\address[label2]{Laboratoire
Particules et Univers de Montpellier, CNRS-IN2P3, 
Case 070, Place Eug\`ene
Bataillon, 34095 - Montpellier, France.}
 \author[label2]{S. Narison\fnref{fn1,fn2}}
  \fntext[fn1]{Madagascar consultant for the Abdus Salam International Centre of Theoretical Physics (ICTP), via Beirut 6,34014 Trieste, Italy .}
  \fntext[fn2]{Corresponding author.}
    \ead{snarison@yahoo.fr}


 \author[label4]{H. Ratsimbarison}
\address[label4]{Institute of High-Energy Physics of Madagascar (iHEP-MAD), University of Antananarivo, 
Madagascar}
\ead{herysedra@yahoo.fr}
\begin{abstract}
\nin
Considering the classical two-point correlators built from (axial)-vector, scalar $\bar qq$ and gluonium currents, we confront results obtained using the SVZ $\oplus$ 1/$q^2$ expansion to the ones from some QCD holographic models in the Euclidian region and with negative dilaton $\Phi_i(z)=- |c_i^2| z^2$. 
We conclude that the presence of the $1/q^2$-term in the SVZ-expansion due to a  tachyonic gluon mass appears naturally in the Minimum Soft Wall (MSW) and the Gauge/String Dual (GSD) models which can also reproduce semi-quantitatively some of the higher dimension condensate contributions appearing in the OPE.  The Hard-Wall model shows a large departure from the SVZ $\oplus$ 1/$q^2$ expansion in the vector, scalar and gluonium channels due to the absence of any power corrections.  The equivalence of the MSW and GSD models is manifest in the vector channel through the relation of the dilaton parameter with the tachyonic gluon mass. For approximately reproducing the phenomenological values of the dimension $d=4,~6$ condensates, the holographic models require a tachyonic gluon mass $(\alpha_s/\pi)\lambda^2\approx -(0.12\sim 0.14)$ GeV$^2$, which is about twice the fitted phenomenological value from $e^+e^-$ data.  The relation of the inverse length parameter $c_i$ to the tachyonic gluon mass also shows that $c_i$ is channel dependent but not universal for a given holographic model. Using the MSW model and $M_\rho=0.78$ GeV as input, we  predict a scalar $\bar qq$ mass $M_S\approx (0.95\sim 1.10)$ GeV and a scalar gluonium mass  $M_G\approx (1.1\sim 1.3)$ GeV.

 \end{abstract}
\end{frontmatter}
\section{Introduction}
The SVZ-sum rules approach \cite{SVZ,SNB} have been used successfully for understanding 
the properties of the lowest mass hadrons in QCD. It is based on a duality between 
the measured spectral function and the semi-perturbative QCD expression using
the Operator Product Expansion (OPE) in terms of the quark and gluon condensates
or their mixing. These condensates are assumed to give a good description of confinement
at moderate energies. More recently \cite{CNZ,ZAK}, it has been also shown that  UV renormalon contributions can be mimics by a $1/q^2$-term induced by a tachyonic gluon
mass while its duality with these large order terms of the PT series has been studied in \cite{SNZ}. \\
On the other hand, QCD holographic models have been recently extensively studied following the ideas in \cite{MALDA,WITTEN} that one can have a duality between the large $N_c$ limit of a maximally ${\cal N}$=4 superconformal $SU(N_c)$ gauge theory in $n$ dimensions and the supergravity limit of a superstring/ M-theory living on a ($n+1$) Anti de Sitter (AdS) space $\otimes$ a compact manifold. Such a gauge/gravity duality has been applied to QCD by mapping it onto gravitational effective theories in 5-dimensions on an AdS background. Different forms of the gauge/string correspondence for studying the hadron spectra in Minkowski space have been proposed in the literature. In the so-called \emph{top-down} approximation, one tries to keep the underlying string structure. In so doing, one usually starts with a theory in 10-dimensions. Its reduction to 5-dimensions in the supergravity limit contains in general  additional higher derivatives which are $1/\sqrt{N_c}$ (or the Regge slope $\alpha'$) corrections. A model of this approach has been discussed in \cite{ANDREEV} and used in
different phenomenological applications in \cite{ANDREEV1,ANDREEV2,ANDREEV3}.
\\
In this paper, we shall consider the so-called Soft- and Hard-Wall models \cite{ERLICH,JUGEAU3} of the \emph{bottom-up} approach and the simplified version of the Gauge/String Duality (GSD) \cite{ANDREEV} used in  \cite{ANDREEV1,ANDREEV2,ANDREEV3}. 
We compare the results obtained for the two-point functions from these holographic models with the SVZ $\oplus$ 1/$q^2$ expansion.
\\
In so doing, we consider the two-point correlators:
\beq
\Pi(q^2)=\imath \int d^4x e^{\imath q\cdot x}\la 0|{\cal T}\{J(x),J^\dagger(0)\} |0\ra,
\eeq
 built from the local (axial-) vector current:
\beq
J^{\mu\,a}_{V/A}(x)=\bar\psi\gamma^\mu(\gamma_5)T^a\psi(x),
\label{eq:vector}
\eeq
or the local scalar quark current:
\beq
J^A_S (x)=\bar\psi T^A\psi(x),
\eeq
where $\psi$ is the light quark field and $T^A=(T^0=1/\sqrt{2n_f},T^a)$ with $T^a\equiv \lambda^a/2$ the $SU(3)_f$ generators ($a=1,\ldots,8$) and the normalization $tr(T^AT^B)=\delta^{AB}/2$.  We shall also consider the gluonium correlator built
from gluonium field strength:
\beq
J_G (x)=\beta(\alpha_s) G_{\mu\nu}^a G^{\mu\nu\,a}(x)~:~~~~\beta(\alpha_s)=\beta_1\ga {\alpha_s\over\pi}\dr+\cdots,
\label{eq:glue}
\eeq 
with $\beta_1=-(1/2)(11-2n_f/3)$ for $n_f$ active flavours.
We shall evaluate the previous two-point correlators using either  the previous holographic models or the SVZ expansion
\cite{SVZ,SNB} including quark and gluon condensates and the $1/q^2$ contribution from tachyonic gluon mass which goes beyond the standard SVZ-expansion \cite{CNZ}. The latter term has been motivated to phenomenologically modelize the UV-renormalon contributions and has been shown in \cite{SNZ} to be dual to large order PT series.
\section{The vector  two-point function $\Pi^{(1)}_{V}(q^2)$}
The two-point correlator has the Lorentz decomposition:
\beq
\Pi^{\mu\nu,\,ab}_{V}(q^2)=-\delta^{ab}\ga g^{\mu\nu}q^2-q^\mu q^\nu\dr \Pi^{(1)}_{V} +\delta^{ab}q^\mu q^\nu \Pi_{V}^{(0)}
\eeq
where the upper indices 1 and 0 refer to the spin 1 (transverse) and spin 0 (longitudinal) hadronic components. 
\subsection*{\b $\Pi^{(1)}_V(q^2)$  using the SVZ expansion $\oplus$ $1/q^2$-term}
\nin
One can evaluate the two-point correlation function using the SVZ expansion $\oplus$ $1/q^2$-term in the Euclidian region.
One obtains to leading order in $\alpha_s$ 
and using the normalization of the current in Eq. (\ref{eq:vector}) \cite{SVZ,SNB,CNZ}:
\bea
\ga{24\pi^2\over N_c}\dr\Pi^{(1)}_{V}(q^2)&=&-\log{-q^2\over \nu^2} +1.05\ga{\alpha_s\over \pi}\dr{\lambda^2\over q^2}\nnb\\
&&+\ga{\pi\over 3}\dr {\la \alpha_s G^2\ra \over q^4}+{896\over 81}\pi^3{\rho\alpha_s\la\bar\psi\psi\ra^2\over q^6}
\nnb\\
&&-\ga 53\log{-q^2\over \nu^2}+{3607\over 122}\dr{\la \alpha_s G^2\ra^2\over 2592q^8},\nnb\\
\label{eq:VVSVZ}
\eea
where we have used the contribution of the $d=8$ condensates within the vacuum saturation 
assumption but including the $1/N_c$ corrections from~\cite{BROAD}. 
$\lambda^2$ is the tachyonic gluon mass with a phenomenological value \cite{SNI,CNZ}:
\beq
\ga {\alpha_s\over\pi}\dr \lambda^2= -(0.06\sim 0.07)~{\rm GeV}^2~,
\label{eq:tach}
\eeq
and with an absolute approximate upper limit:
\beq
\ga {\alpha_s\over\pi}\dr \vert\lambda^2\vert\leq (0.12\sim 0.14)~{\rm  GeV}^2~.
\label{eq:tachbound}
\eeq
\subsection*{\b $\Pi^{(1)}_{V}(q^2)$ from the MSW model with a positive dilaton field}
\nin
The vector two-point function in the Minimal Soft Wall (MSW) model has been studied in \cite{JUGEAU3,ZUO}, where the case of  a Minkowskian metric and a positive  background dilaton field  $\Phi_v(z) =c_v^2z^2$ with $c_v^2>0$ has been considered. The models are characterized by the metric:
\beq
ds^2 = e^{2A(z)}\left(\eta_{\mu\nu}dx^{\mu}dx^{\nu} + dz^2\right) \equiv g_{MN}dx^Mdx^N ~,
\eeq
where $A(z)=-\ln(z/R)$ for pure AdS$_5$ with radius $R$; $\eta_{\mu\nu}\equiv g_{\mu\nu}$ is the 4D Minkowski metric tensor with signature (-+++). The bulk action for a gauge vector field over the bulk-volume ${\cal V}$ is given by \cite{JUGEAU3}:
\beq
	S_{5d}^{eff}[V] = -\frac{1}{k}\int d{\cal V}e^{-\Phi(z)}\frac{1}{2g_5^2}tr\,F_{(V)}^2~,
\label{eq:action}	
\eeq
where $k$ is the 5-dimensional Newton constant and $F_{(V)\,MN}^a=\partial_{M}V_N^a-\partial_NV_M^a$ is the abelian part of the field strength tensor. The corresponding equation of motion reads:
\beq
\partial_M(\sqrt{-g}e^{-\Phi(z)}F_{(V)}^{MN\,a})= 0~,
\eeq
which, in terms of the $4d$ Fourier-transformed field $\tilde{V}_M^a(q,z)$ of $V_M^a(x,z)$,  becomes in the axial-like gauge $V_z^a=0$:
\beq
\partial_z\left(\frac{R}{z}e^{-\Phi(z)}\partial_z\tilde{V}^{\mu\,a}(q,z)\right)-q^2\frac{R}{z}e^{-\Phi(z)}\tilde{V}^{\mu\,a}(q,z)=0\label{firstsoluce}
\eeq
and:
\beq
q_{\mu}\tilde{V}^{\mu\,a}(q,z)=0~.
\eeq
The equation of motion of the bulk-to-boundary propagator reads:
\beq
\left[\partial^2_z - \left(2c_v^2z + \frac{1}{z}\right)\partial_z - q^2\right]V\ga\frac{q^2}{c_v^2},c_v^2z^2\dr = 0~,
   \label{qqbEbeom}
\eeq
with the general solution ($z>0$):
\bea
V\ga\frac{q^2}{c_v^2},c_v^2z^2\dr&=&A U\left(\frac{q^2}{4c_v^2}\,;\,0\,;\,c_v^2z^2\right)\nnb\\
 &&+B(q^2)\,c_v^2z^2\,{_1F_1}\left(\frac{q^2}{4c_v^2}+1\,;\,2\,;\,c_v^2z^2\right)
 \label{eq:B}
 \eea
 where 
\beq
 A=\Gamma\ga 1+{q^2\over 4c_v^2}\dr~
 \label{eq:digamma}
 \eeq
 and $B$ are integration constants with respect to $z$; $U$ is the Tricomi confluent hypergeometric function and $_1F_1$ is the Kummer confluent hypergeometric function. In the UV limit $z\to 0$, this solution reads:
 \bea
V\ga\frac{q^2}{c_v^2},c_v^2z^2\dr&=&1+\frac{q^2z^2}{4}\Big{[}\ln(c_v^2z^2)+\psi\left(\frac{q^2}{4c_v^2}+1\right)\nnb\\
&&+2\gamma_E-1\Big{]}
+B(q^2) c_v^2z^2+O(z^4)~,
\label{gensolEqqb}
\eea
where the digamma function $\psi(a)$ has an infinite set of simple poles located at $a=-n$ ($n\in\mathbb{N}$) with residues -1 and
\beq
 \psi(a)=\ln{a}-{1\over 2a}-\sum_{k=1}^\infty{{B_{2k}\over 2k}{1\over a^{2k}}}~,
 \eeq
 where $B_2=1/6$, $B_4=-1/30$,... are Bernoulli numbers. 
In the Fourier $q$-space, the two-point correlator is obtained by means of twice the functional derivative of the critical bulk action with respect to the boundary field $\tilde{V}_0$:
\bea
&&-\delta^{ab}\left(\eta^{\mu\nu}q^2-q^{\mu}q^{\nu}\right)\Pi_V^{(1)}(q^2)=\nnb\\
&&-\imath(2\pi)^4\int d^4 k\frac{\delta^2}{\delta\tilde{V}^a_{0\,\mu}(q)\tilde{V}^{b}_{0\,\nu}(k)}e^{\imath {S_{5d}^{eff}}[\tilde{V}_{crit}]}\big{|}_{\tilde{V}_0=0}~,
\eea
from which one can deduce:
\bea
&\Pi^{(1)}_{V}(q^2)=-\frac{1}{q^2}\frac{R}{kg_5^2}V\ga\frac{q^2}{c_v^2},c_v^2z^2\dr\frac{e^{-\Phi(z)}}{z}\partial_z 
V\ga\frac{q^2}{c_v^2},c_v^2z^2\dr\Big{|}_{z=\frac{1}{\nu}\rightarrow 0}
\label{eq:prop}
\eea
where the critical field $\tilde{V}_{crit}(q,z)$, also solution of Eq. (\ref{qqbEbeom}), is related to its restriction $\tilde{V}_0(q)$ on the boundary space by means of the bulk-to-boundary propagator $V(\frac{q^2}{c_v^2},c_v^2z^2)$:
\begin{eqnarray}
\tilde{V}^a_{\mu\,crit}(q,z) =V\ga\frac{q^2}{c_v^2},c_v^2z^2\dr\tilde{V}^a_{0\,\mu}(q)~. 
\end{eqnarray}
Using the solution in Eq.(\ref{eq:prop}), one can deduce 
  in the Minkowski space \cite{JUGEAU3}:
\beq
\ga{kg_5^2\over R}\dr\Pi^{(1)}_{V}=-\frac{1}{2}\log{q^2\over \nu^2} -{c_v^2\over q^2}+{2\over 3}{c_v^4\over q^4}-{16\over 15}{c_v^8\over q^8}+{\cal O}\ga\frac{1}{q^{14}}\dr\\
\label{eq:VV}
\eeq
where we have dropped constant terms which are physically irrelevant; $\nu$ is the usual subtraction constant. It is important to notice that due to an ``accident"  numerical cancellation among different terms, the coefficients of $c_v^6$, $c_v^{10}$ and $c_v^{12}$ vanish in the expansion. 
By comparing this result obtained in the Minkowskian region with the SVZ one in the Euclidian space, 
one should first Wick rotate the action in Eq. (\ref{eq:action}) such that $q_M^2>0$  becomes $-q_E^2>0$ (in the signature (-++++) and where the subscripts M and E denote respectively Minkowski and Euclidian \footnote{In the remaining part of the paper after Eq. (\ref{eq:substit}), we shall omit these indices for simplifying the notation.}) and change $q_M^2$ to $-q_E^2$:
\beq
q_M^2>0 \to -q_E^2> 0~.
\label{eq:substit}
\eeq

\subsection*{\b $\Pi^{(1)}_{V}(q^2)$ from an extension of the MSW model}
\nin
The absence of the $1/q^6$ term in the MSW model due to some miraculous cancellation of its coefficients shows its departure from the SVZ-expansion. One can try to improve the model by 
adding the contribution of the large $q^2$ limit $\tilde B$ of the $B$-term in Eq. (\ref{eq:B}). In this way, one obtains the additionnal contribution:
\beq
\left(\frac{kg_5^2}{R}\right)\delta\Pi_V^B(q^2)=-{2\over q^2}\tilde{B}(q^2)c_v^2~
\eeq
where $\tilde{B}(q^2)$ is the asymptotic form of $B(q^2)$ in the large $q^2$ limit. The presence of this term is however unlikely since this solution gives a divergent action $S_{5d}^{eff}$ when $z\rightarrow+\infty$ contrary to the prescription in MSW model where the solutions correspond to a regular action.   Keeping this solution in $B$ would be similar to another choice of regularization prescription. In this way, one can obtain new terms:
\beq
\tilde B(q^2)= \tilde B_0+\tilde B_2{c^2_v\over q^{2}}+ \tilde B_4{c^4_v\over q^{4}}+ \tilde B_6{c^6_v\over q^{6}}+\cdots,
\label{eq:Bt}
\eeq
 where $\tilde B_i$ are arbitrary coefficients such that the modified model looses its predictivity and becomes less interesting. 
\subsection*{\b $\Pi^{(1)}_{V}(q^2)$ from the MSW model with a negative dilaton field}
\nin
One can do an analysis similar to the case of the positive dilaton field \cite{JUG4}. Defining the negative
dilaton field by:
\beq
\Phi_v(z)=-|c_v^2|z^2~,
\eeq
the corresponding bulk propagator reads:
\bea
V_{<0}&=&V(c_v^2\lr |c_v|^2)e^{-|c_v|^2z^2}\nnb\\
&=&V(z\to 0)-|c_v|^2z^2+{\cal O}(z^3)~,\label{poleneg}
\eea
where $V$ is the bulk propagator for a positive dilaton field defined in Eq. (\ref{eq:B}). Using Eq. (\ref{eq:prop}) for $V_{<0}$, one can deduce in the Minkowski space:
\beq
\Pi^{(1)}_{V}(q^2,-|c^2_v|)=\Pi^{(1)}_{V}(q^2,c^2_v>0)+{2|c_v^2|\over q^2}~,
\eeq
where there is an additional term (interpreted as a massless unphysical pole in the literature) compared with the expression of $\Pi^{(1)}_{V}$ for $c^2_v>0$. 
Expanding the previous expression for $q^2\to +\infty$, one obtains:
\bea
\ga{kg_5^2\over R}\dr\Pi^{(1)}_{V}&=&-\frac{1}{2}\log{q^2\over \nu^2} -{|c_v|^2\over q^2}+{2\over 3}{|c_v^4|\over q^4}\nnb\\
&&-{16\over 15}{|c_v^8|\over q^8}+
{\cal O}\ga\frac{1}{q^{14}}\dr+\cdots+{2|c_v^2|\over q^2}~,
\label{eq:VV2}
\eea
where the additionnal term proper to $c_v^2<0$ changes the sign of the total contribution of the $1/q^2$-term. Translating the result to the Euclidian region by changing $q^2>0$ to $q^2<0$, the final result
becomes the same as in the case of a positive dilaton field in Eq. (\ref{eq:VV}) when translated to the Euclidian region. From this exercise, we show that the expression of $\Pi^{(1)}_{V}$ for $\Phi_v(z)=\pm |c_v^2|z^2$ remains the same thanks to the existence of this additional $1/q^2$-term, in the case of a negative $c_v^2$, where its physical interpretation remains to be clarified. 
\subsection*{\b $\Pi^{(1)}_{V}(q^2)$ from Hard Wall (HW) model  }
\nin
Evaluating the vector correlator in the HW model, one obtains~\cite{KIRKUN}:
\beq
\ga{kg_5^2\over R}\dr\Pi^{(1)}_{V}(q^2)=-\frac{1}{2}\log{q^2\over \nu^2} -\gamma_E+\ln2~,
\eeq
where the perturbative part is the same as in the MSW model which is natural because the two models only differ in the IR region. However, it is rewarding that in the HW model, $\Pi^{(1)}_{V}$ has no power corrections at all indicating its large departure from the SVZ-expansion. 
\subsection*{\b $\Pi^{(1)}_{V}(q^2)$ from a gauge/string dual (GSD) model}
\nin
Using a gauge/string dual (GSD) model \cite{ANDREEV}, $\Pi^{(1)}_{V}$ has been evaluated until the $1/q^2$ term in \cite{ANDREEV1} in the background given by:
\beq\label{GSD modelmetric}
ds^2 = {R^2\over z^2}e^{-\frac{1}{2}c_A z^2}\left(\eta_{\mu\nu}dx^{\mu}dx^{\nu} + dz^2\right) \equiv g_{MN}dx^Mdx^N ~,
\eeq
with the result:
\beq
\Pi^{(1)}_{V}(q^2)\sim -\log{q^2\over \nu^2}-{3c_A\over q^2}~.
\eeq
The coefficient of the $1/q^2$ term can be compared with the one from the MSW model  in Eq. (\ref{eq:VV}) from which one can deduce:
\beq
2c_v^2=3c_A~.
\label{eq:andreev}
\eeq
By using a Weyl tranformation, one can show that the MSW model and GSD model are indeed equivalent in the vector channel taking into account the constraint in Eq. (\ref{eq:andreev}) though \cite{ANDREEV1} derives a slightly different constraint $2c_v^2=c_A/2$ which is due to the fact that he started from a string-like action in the background (\ref{GSD modelmetric}) which differs from the one in Eq. (\ref{eq:action}). 

\subsection*{\b SVZ $\oplus~ 1/q^2$-expansion versus some QCD holographic models}
\nin
-- Matching the SVZ $\oplus~ 1/q^2$-expansion result in the Euclidian region in Eq. (\ref{eq:VVSVZ}), with the previous results from different holographic models, one can remark that the HW model does not  present any power corrections at all indicating its large departure from the SVZ-idea. \\
-- From the previous analysis, one can notice that the quadratic correction introduced by \cite{CNZ} appears naturally in the MSW model and GSD model. This quadratic correction could be a remnant of stringy effect in QCD as suggested by Zakharov.  Confronting the expressions in Eqs. (\ref{eq:VVSVZ}) and (\ref{eq:VV})  by doing the substitution in Eq. (\ref{eq:substit}), leads to:
\beq
{kg_5^2\over R}= {12\pi^2\over N_c}~~~~~\rm{and}~~~~~2c_v^2=1.05\ga{\alpha_s\over \pi}\dr\lambda^2~.
\eeq
Using the previous value of $\lambda^2$ in Eq. (\ref{eq:tach}), one can deduce:
\beq
c_v^2=-(0.03\sim 0.04)~{\rm GeV}^2 <0~,
\eeq
where the sign of $c_v^2$ is consistent with a tachyonic gluon. \\
-- Using this result, one can predict for the $d=4$ condensate:
\beq
\la \alpha_s G^2\ra  = {4\over \pi}c_v^4\simeq (1\sim 2)\times 10^{-3}~{\rm GeV}^4~,
\eeq
which is relatively small compared to the phenomenological value \cite{SNTAU,LNT,SNI,fesr,YNDU,SNHeavy,BELL,SNG2,SNH10}:
\beq
\la \alpha_s G^2\ra= (0.07\pm 0.02)~{\rm GeV}^4~,
\label{eq:g2}
\eeq
and to the original SVZ estimate $\la \alpha_s G^2\ra\approx 0.04$ GeV$^4$ \cite{SVZ} which has been recovered in \cite{ANDREEV2} from a direct evaluation of the Wilson loop in the GSD model. An analogous
underestimate occurs for the $d=8$ condensate obtained using a vacuum saturation assumption but including the $1/N_c$ corrections. In this $d=8$ case, one can also notice that the MSW model does not reproduce the $\log{-q^2/\nu^2}$-term appearing in the OPE. \\
-- Inversely, we can instead use the value of $\la \alpha_s G^2\ra$
in Eq. (\ref{eq:g2}) to fix $|c_v^2|$. We obtain in this way:
\beq
|c_v^2|=0.23~{\rm GeV}^2~\lrar~\ga {\alpha_s\over\pi}\dr |\lambda^2|= 0.45~{\rm GeV}^2~.
\label{eq:cv1}
\eeq
Instead, one can also use the relation between
$|c_v^2|$ and the $\rho$-meson spectrum \cite{JUGEAU3,JUG4,POMAROL}:
\beq
M_{\rho_n}^2=4\vert c_v^2 \vert (n+1)~,
\eeq
coming from the pole of the digamma function $\psi\ga 1+{q^2/ 4|c_v|^2}\dr$ appearing in Eq. (\ref{poleneg}) when $q^2=-M_{\rho_n}^2$.
Demanding that the MSW model reproduces the lowest ground state  $\rho$-meson ($n=0$), one can deduce:
\beq
|c_v^2|\simeq 0.15~{\rm GeV}^2 \lrar \ga {\alpha_s\over\pi}\dr |\lambda^2|= 0.30~{\rm GeV}^2~.
\label{eq:cv2}
\eeq
The absolute values in Eqs. (\ref{eq:cv1}) and (\ref{eq:cv2}) are much larger than the phenomenological upper value of about $(0.12\sim 0.14)$ GeV$^2$ obtained in~\cite{SNI}. 
Within the crude approximation where these numbers have been derived, we consider
that the agreement with the phenomenological fit is remarkable. It is also important to notice that the matching of the MSW model results in the Euclidian region with the SVZ-expansion is essential for fixing the sign of $c_v^2$ which cannot be done from the alone fit of the spectrum in the Minkowski
space due to the fact that the bulk potential from the Schr\"odinger-type equation behaves like~$c_v^4$.\\
-- As already mentioned earlier, the $d=6$ condensate is absent in the MSW model due to some numerical cancellation of its coefficient. The extension of the MSW model can cure this disease by the introduction of a non-zero integration constant $\tilde B$ like in Eq. (\ref{eq:Bt}). However, within this framework, the MSW model looses its predictivity due to the arbitrariness of $\tilde B$ and to the lack of convergence of the corresponding effective action $S_{5d}^{eff}$ when $z\rightarrow+\infty$.\\
-- A result similar to the one in the MSW model has been obtained by \cite{ANDREEV1} in the GSD model. 

\section{The $\Pi^{(1)}_{LR}(q^2)\equiv \Pi^{(1)}_{V}(q^2)-\Pi^{(1)}_{A}(q^2)$ correlator}
$\Pi_{LR}$ controls the $SU(n)_L\times SU(n)_R$ global chiral symmetry~\cite{SNB} and controls the assumption of the superconvergence of the 1$^{st}$ (which involves the sum of the spin 1 and 0 contributions) and the 2$^{nd}$ (which involves the spin 1 or 0 contribution) Weinberg sum rules for $q^2\to \infty$ \cite{WEINBERG}. Solving these sum rules in the chiral limit $m_q=0$ and the current algebra relation in Ref. \cite{SUZUKI}, Weinberg has derived the well-known relation:
\beq
M_{A_1}\simeq \sqrt{2}M_\rho~.
\eeq
 It has been shown within QCD \cite{FNR} that the perturbative (PT) contribution due to the light quark masses breaks the 1$^{st}$ Weinberg sum rule  to order $\alpha_s m_q^2$ while the 2$^{nd}$ sum rule is broken to leading order $m_q^2 \log m^2_q$ of PT.  
\subsection*{\b  $\Pi^{(1)}_{LR}(q^2)$ in the SVZ $\oplus$ $1/q^2$ expansion}
\nin
In the chiral limit $m_q=0$, and using the SVZ expansion,  one also finds  a cancellation of contributions of terms of dimension $d\leq 4$ due to $m_q\la\bar qq\ra$ and the flavour 
independent $1/q^2$ and $\la\alpha_s G^2\ra$ contributions. 
One obtains \cite{SVZ,SNB,SNWSR,KNECHT}:
\bea
q^6\Pi^{(1)}_{LR}&=&{8\pi} \rho \alpha_s\la \bar uu\ra ^2~,\nnb\\
q^6\Pi^{(1+0)}_{LR}&=&{32\pi\over 9} \rho \alpha_s\la \bar uu\ra ^2~,
\label{eq:lrsvz}
\eea
where $\rho\simeq 2$ indicates a deviation from vacuum saturation. An extraction of the four-quark condensate from $e^+e^-$ and $\tau$-decay data and the light baryon spectra gives \cite{SNTAU,LNT,JAMI2}:
\beq
\rho\alpha_s\la \bar uu\ra^2\simeq (4.5\pm 0.3)\times 10^{-3}~{\rm GeV}^6~.
\label{eq:4quark}
\eeq
\subsection*{\b  $\Pi^{(1)}_{LR}(q^2)$ in the HW model}
\nin
 One obtains in the Hard-Wall model~\cite{KIRKUN}:
\beq
q^6\Pi^{(1)}_{LR}={4\over 5}{\pi^2\over N_c}\la \bar uu\ra^2~,
\label{eq:hard}
\eeq
where all contributions come from $\Pi^{(1)}_{A}$ due to the fact that $\Pi^{(1)}_{V}$ has no power corrections in this model. However, this feature is completely different from the SVZ-expansion. 
\subsection*{\b  $\Pi^{(1)}_{LR}(q^2)$ in the MSW model}
\nin
In the MSW model, the linear equation of motion of the  axial-vector bulk fields (which describe holographically the axial resonances) reads as:
\bea
&&\partial_z\left(\frac{R}{z}e^{-\Phi(z)}\partial_z A_{\perp}\right)-q^2\frac{R}{z}e^{-\Phi(z)}A_{\perp}\nnb\\
&&-\frac{R}{z}e^{-\Phi(z)}\frac{g_5^2R^2v(z)^2}{z^2}A_{\perp}=0~.
\eea
$v(z)$ is the function which describes the chiral symmetry breaking. Typically, $v(z)$ has two contributions: one which depends on the mass $m_q$ of the light quark while the other depends on the chiral condensate $\langle\bar{q}q\rangle$. 
As a result, when $m_q=0$ and keeping only the contributions of condensates $d\leq 4$, $v(z)^2$ vanishes and the vector and the axial sectors are completely equivalent [see Eq. (\ref{firstsoluce})]. 
In the MSW model, the expression of $\Pi_{LR}$ similar to the one in Eq. (\ref{eq:hard}) has been derived in \cite{COLAN} from a relation of $\Pi_{LR}$ with the axial-vector-vector (AVV) vertex following the evaluation of this vertex in \cite{KNECHT2}.  We plan to come back to this analysis
in a future publication. 
\subsection*{\b SVZ $\oplus~ 1/q^2$-expansion versus holographic models}
\nin
One can compare the previous expressions (\ref{eq:lrsvz}) and (\ref{eq:hard}) in the large $N_c$ limit. Using:
\beq
\ga\alpha_s\over\pi\dr={1\over -\beta_1}\approx {6\over 11N_c}~,
\eeq
one can notice that the previous  expressions exhibit qualitatively the same $\pi^2$ and large $N_c$ behaviours\,\footnote{The proper behaviour in the large-$N_c$ limit of the HW model has been e.g. discussed in \cite{FJ}.} but with different numerical values.
\section{The scalar  two-point function $\Pi_{S}(q^2)$}
\subsection*{\b $\Pi_S(q^2)$  using the SVZ expansion $\oplus$ 1/$q^2$-term}
\nin
We pursue our investigation by evaluating the two-point function of the scalar current using the SVZ $\oplus$ $1/q^2$ expansion. We shall
work in the chiral limit $m_q=0$. Detailed derivations of the expression have been extensively
discussed in the literature. To leading order in $\alpha_s$, the result is  \cite{SVZ,SNB}:
\bea
2\ga{ 8\pi^2\over N_c}\dr\Pi_S(q^2)&=&-\Big{[}q^2-4\ga {\alpha_s\over\pi}\dr \lambda^2\Big{]}\log{-q^2\over \nu^2} \nnb\\
 &&-{\pi\over 3}{\la \alpha_s G^2\ra \over q^2}-{176\pi\over 27}{\rho\alpha_s\la \bar uu\ra^2\over q^4}-\nnb\\
&& \ga \log{-q^2\over \nu^2}-{3677\over 192}\dr{\la \alpha_s G^2\ra^2\over 1728q^6}~,\label{scalarmesonSVZ}
 \eea
where we have used the result obtained by \cite{BROAD} for the $d=8$ condensate contributions.
\subsection*{\b $\Pi_{S}(q^2)$ from the MSW model}
\nin
The expression of this correlator is known in the literature for the case of a positive dilaton\,\footnote{One should notice that $\Pi_{S}(q^2)$ does not have any short distance  $1/q^{2n}$ power corrections in the HW model.}. It reads in the Euclidean space \cite{JUGEAU,HERY}
\bea
\ga {k\over R}\dr\Pi_S(q^2)&=&-(q^2-2c_s^2)\log\ga {-q^2\over \nu^2}\dr -{2\over 3}{c_s^4\over q^2}\nnb\\ 
&&+{4\over 3}{c_s^6\over q^4}+{28\over 15}{c_s^8\over q^6}-{56\over 15}{c_s^{10}\over q^8}+\cdots~\,,
\eea
where we have dropped constant terms which are physically irrelevant; $\nu$ is the usual subtraction constant. We do the substitution in Eq. (\ref{eq:substit}) for transfering this result into the Euclidian region, where one can also notice that the result remains the same for a negative dilaton.

\subsection*{\b SVZ $\oplus~ 1/q^2$-expansion versus MSW model}
\nin
Matching the result from the two different approaches in the Euclidian region, one can deduce:
\beq
{k\over R}={16\pi^2\over N_c},~~~c_s^2=2\ga{\alpha_s\over\pi}\dr \lambda^2~,
\eeq
where $c_s^2$ is {\it negative} due to the tachyonic value of the gluon mass. Using this relation and the phenomenological value of $\alpha_s\lambda^2$ in Eq. (\ref{eq:tach}), one can predict:
\beq
\la \alpha_s G^2\ra={2\over \pi}c_s^4\approx (0.02-0.04)~{\rm GeV^2}~.
\eeq
For the $d=6$ contribution, the negative value of $c_s^2$ is crucial for matching the
two approaches. One can deduce:
\beq
\rho\alpha_s\la \bar qq\ra^2=-{9\over 44\pi}c_s^6\approx 1.1 \times 10^{-3}~,
\eeq
compared with the phenomenological estimate in Eq. (\ref{eq:4quark}). 
One can notice that the AdS approach tends to underestimate the value of the $d=4$ and $d=6$ condensates compared with the most recent phenomenological values. 
Alternatively, if one takes seriously the value of $\la \alpha_s G^2\ra=0.07$ GeV$^4$ from more recent phenomenology, one can predict:
\beq
|c_s^2|\approx 0.34 ~{\rm GeV^2}~,
\eeq
while a deviation by a factor 2 above the vacuum saturation of the four-quark condensates leads to:
\beq
c_s^2\approx -0.30 ~{\rm GeV^2}~,
\eeq
where the sign has been fixed by the matching of the $d=6$ condensate contribution which  requires again a tachyonic gluon mass.
Therefore, our analysis shows that matching the MSW model approach with the SVZ expansion requires a value of the tachyonic gluon mass to be in the range:
\beq
\ga{\alpha_s\over\pi}\dr \lambda^2\approx -(0.15\sim 0.17)~{\rm GeV}^2~,
\label{eq:lambda2}
\eeq
where the absolute value is slightly higher but consistent with phenomenological determinations. Moreover, it is more important to notice that the tachyonic origin of the gluon mass is necessary for matching the MSW model  and the SVZ $\oplus$ $1/q^2$-expansion in the Euclidian region in order to have the correct signs of the $d=2$ and $d=6$ contributions.
\section{The scalar gluonium two-point function}
\subsection*{\b $\Pi_G(q^2)$  from SVZ  $\oplus$ 1/$q^2$ expansion}
\nin
 Using SVZ-expansion for large $q^2$ in the Euclidian region and including the quadratic correction introduced by CNZ \cite{CNZ}
 for modelling UV renormalons and/or large order terms of the PT series, the expression
of the two-point function reads for $q^2<0$ and to leading order in $\alpha_s$ \cite{NSVZ,SNG}:
 \bea
{1\over 2} \ga {\pi\over \beta(\alpha_s)}\dr^2 \Pi_G(q^2)&=&-\Big{[}(q^2)^2-\lambda^2q^2\Big{]}\log {-q^2\over \nu^2} \nnb\\
&&+ {2\pi^2\over\alpha_s} \la \alpha_s G^2\ra-{\pi\over\alpha_s}{\la g_s^3 G^3\ra\over q^2}\nnb\\
&&+{9\over 8}{\pi^3\over\alpha_s}{\la \alpha_s G^2\ra^2\over q^4}~,
\eea
where we have used a vacuum saturation including the $1/N_c$ corrections for the estimate of the $d=8$ condensate contributions;
$\beta(\alpha_s)$ is the usual $\beta$-function defined in Eq. (\ref{eq:glue}) and $\lambda^2$ is the square of the tachyonic gluon mass. 
\subsection*{\b $\Pi_G(q^2)$ from MSW model}
\nin
The two-point correlator  has been evaluated in the MSW model
by \cite{JUGEAU2,FORKEL1} in the Minkowski space and for a positive dilaton. Using the substitution in Eq. (\ref{eq:substit}), we can transform  the previous result to the Euclidian space ($q^2<0$):
\bea
\ga{8k\over R^3}\dr \Pi_G(q^2)&=&-\Big{[}(q^2)^2-4c_g^2q^2\Big{]}\log {-q^2\over \nu^2}\nnb\\ &&+{4\over 3}c_g^4\big{[} 12\tilde B(q^2/c_g^2) -5\big{]}-{16\over 3}{c_g^6\over q^2}\nnb\\
&&-{32\over 15}{c_g^8\over q^4}+\cdots~,
\eea
where one can also notice that the result is the same for a negative dilaton. 
$\tilde B(q^2)$ is a function of $1/q^2$ like in Eq. (\ref{eq:Bt}) but constant with respect to the variable $z$.  In the MSW model, one takes $\tilde B$=0. In the HW model, one again finds that $\Pi_G(q^2)$ does not have any power corrections like in the vector and scalar $\bar qq$ channels.
\subsection*{\b SVZ $\oplus~ 1/q^2$-expansion versus MSW model}
\nin
We compare the AdS/QCD expression in the Euclidian space with the one from the SVZ-expansion. \\
-- One obtains by matching the leading order PT expressions:
\beq
{R^3\over 8k}=2\ga{\beta(\alpha_s)\over\pi}\dr^2~.
\eeq
where $\beta(\alpha_s)$ is  defined in Eq. (\ref{eq:glue}).\\
-- Equating the quadratic terms, one obtains:
\beq
4c_g^2= \lambda^2~.
\label{eq:cprim}
\eeq
-- Equating the $d\geq 4$ terms, one finds in the MSW model that all of these terms have opposite signs
with the ones in the SVZ-expansion. 
Such problems can be evaded by assuming that $\tilde B\not=0$ and by working within the expansion in Eq. (\ref{eq:Bt}) like done in \cite{JUGEAU}. However, the approach becomes obviously less predictive
and would imply modifications in the Soft Wall recipe.
\section{Tachyonic gluon mass and hadron mass scale hierarchy }
\subsection*{\b Tachyonic gluon and hadron spectra}
\nin
The parameter $c^2_i>0$ which is the inverse of a squared length and which parametrizes the strength of a positive dilaton field $\Phi_i(z)= c^2_iz^2$
is assumed in the existing literature to be universal for a given model:
\beq
c_v^2=c_s^2=c_g^2~.
\label{eq:coeff}
\eeq
Therefore, from the mass relation for the $n^{th}$-radial excitation in the Minkowski space and 
\cite{JUGEAU3,JUGEAU,JUGEAU2}:
\bea
M^2_{\rho_n}&=&4c^2_v\ga n+1\dr~,\nnb\\
M^2_{S_n}&=&4c^2_s\ga n+{3\over 2}\dr~,\nnb\\
M^2_{G_n}&=&4c^2_g\ga n+2\dr~,
\label{eq:mass}
\eea
it is usual to deduce the ratios of the ground state masses ($n=0$):
\beq
M^2_S\approx {3\over 2}M^2_\rho~,~~~~~~M^2_G\approx 2 M^2_\rho~.
\label{eq:cpositif}
\eeq
One can also write the analogue of the previous results for the case of a negative dilaton field $\Phi_i(z)=- |c^2_i|z^2$. One obtains \cite{JUG4}:
\bea
M^2_{\rho_n}&=&4|c^2_v|\ga n+1\dr~,\nnb\\
M^2_{S_n}&=&4|c^2_s|\ga n+{1\over 2}\dr~,\nnb\\
M^2_{G_n}&=&4|c^2_g|\ga n+1\dr~.
\label{eq:mass}
\eea
If one assumes the equality in Eq. (\ref{eq:coeff}), one would obtain:
\beq
M^2_S\approx {M^2_\rho\over 2}~,~~~~~~M^2_G\approx  M^2_\rho~.
\label{eq:mrho1}
\eeq
$M_S$ is relatively low compared to the QCD spectral sum rules (QSSR) results for a $\bar qq$ meson mass $M_{S_2}\approx 1$ GeV \cite{SNB}, while the one of $M_G$ is consistent with the one from a subtracted gluonium sum rule: $M_{G_1}\approx 1$ GeV  \cite{SNG,VENEZIA,SNG0}. 
\\
However, from our previous analysis in this paper and the relations between $c_i$ and the tachyonic gluon mass, one obtains instead of Eq. (\ref{eq:coeff}) the new relation for a negative dilaton:
\beq
2 |c_v^2|\approx {|c_s^2|\over 2}\approx 4\ga {\alpha_s\over \pi}\dr |c_g^2|\approx \ga {\alpha_s\over \pi}\dr |\lambda^2|~,
\label{eq:coeff2}
\eeq
indicating that $c_i$ is not necessary universal contrary to the tachyonic gluon mass $\lambda^2$.
As shown in previous analysis, the value of the tachyonic gluon mass  from phenomenological fits in Eq. (\ref{eq:tach}) used in the holographic models leads to an underestimate of the QCD vacuum condensates. The value of $\lambda^2$ which leads to the correct phenomenological values of the
gluon  $\la \alpha_s G^2\ra$ and four-quark $\la \bar uu\ra^2$ condensates ranges from the
ones in Eqs. (\ref{eq:cv1}), (\ref{eq:cv2}) and (\ref{eq:lambda2}):
\beq
\ga {\alpha_s\over \pi}\dr\lambda^2\approx -(0.15\sim 0.45)~{\rm GeV}^2~,
\eeq
which is at the limit of the upper value obtained from phenomenological fit in Eq. (\ref{eq:tachbound}). In the following discussion, we shall use the compromise value:
\beq
\ga {\alpha_s\over \pi}\dr\lambda^2\approx -(0.12\sim 0.14)~{\rm GeV}^2~,
\label{eq:lambdabound}
\eeq
from the upper value in Eq. (\ref{eq:tachbound}). From Eqs. (\ref{eq:mass}) and (\ref{eq:coeff2}), one
can deduce in units of GeV:
\beq
M_\rho\approx 0.5~,
\eeq
which implies:
\beq
 {M_S}\approx \sqrt{2}M_\rho\approx  0.7~,~~~~ M_G\approx \sqrt{\pi/2\alpha_s}M_\rho \approx 0.8~,
\label{eq:mrho2}
\eeq
where we have used  $(\alpha_s/\pi)\approx -1/\beta_1\simeq 2/11$ in the large $N_c$-limit.
Taking as a final estimate the range of values obtained in Eq. (\ref{eq:cpositif}) from a positive and  in Eq.  (\ref{eq:mrho2})  from a negative dilaton, one can deduce:
\beq
 {M_S^2\over M_\rho^2}\approx {1.5}\sim 2.0~,~~~~{M^2_G\over M_\rho^2}\approx 2\sim {\pi\over 2\alpha_s}\approx 2.0\sim 2.8~.
\eeq
Assuming that the systematic uncertainties of the approach tend to cancel for the ratios of masses, we shall use the experimental value of $M_\rho=0.778$ GeV in order to deduce the final conservative results:
\beq
M_S\approx (0.95\sim 1.1)~{\rm GeV}~,~~~~~~M_G\approx (1.1\sim 1.3)~{\rm GeV}~.
\eeq
The masses of the vector and scalar mesons agree within 20\% with the experimental value [$\rho(0.78)$ and $a_0(0.98)$] which is the expected accuracy of these leading large $N_c$-limit holographic models. The predicted mass of the gluonium is in the range predicted by the subtracted  ($M_{G_1} \simeq 1$ GeV) and unsubstracted ($M_{G_2} \simeq 1.5$ GeV) QCD gluonium spectral sum rules \cite{SNG,VENEZIA,SNG0} where the presence of two gluonia states are necessary for a consistency of the two sum rules {\large [}$G_1$ is expected to couple strongly to pion pairs and might be identified to the $\sigma(500)$ \cite{VENEZIA,MINK,SNG} as indicated from recent fits of $\pi\pi$ and $\gamma\gamma$ scatterings data \cite{MENES}, while $G_2$ is the ``real glueball" which decays into the $\eta'\eta,...U(1)$ channels{\large ]}. The value of $M_G$ is consistent with the one from instanton sum rule \cite{FORKEL1} and with the lattice results of about 1 GeV including dynamical quarks \cite{HART} but lower than the quenched result of 1.6 GeV \cite{LATT}\,\footnote{For a recent review on the glueball status, see e.g. \cite{OCHS}. Some other attempts to estimate the glueball masses in some holographic models can be found in \cite{FORKEL2}.}. 

\subsection*{\b Hadronic mass scale}
\nin
Like in the QCD spectral sum rule analysis \cite{CNZ}, one can also use $c_i$ for fixing the scale of the analyzed hadronic channel which is not necessary universal\,\footnote{These non-universality of $c_i$ can be also signaled by the difference of PT radiative corrections in each channel which has also been used in \cite{KATAEV} for fixing the hadronic scale hierarchy.}. Following \cite{CNZ},
one can deduce the hadronic scale from Eq. (\ref{eq:coeff2}):
\bea
\Lambda^2_S&\approx& 2|c_s|^2\approx 4\Lambda^2_{\rho}\approx 4.2\ga\alpha_s/ \pi\dr|\lambda^2|~,
\nnb\\
\Lambda_G^2&\approx& 4|c_g|^2\approx|\lambda^2|\approx
5.5\Lambda_\rho^2~,
\eea
where we have used $(\alpha_s/\pi)\approx 2/11$. The sizes of these scales are similar to the optimization scale appearing in the QCD spectral analysis of each channels
but they are not necessary the hadron masses.  Our results indicate the scale hierarchy:
\beq
\Lambda_{\rho}< \Lambda_S< \Lambda_G~,
\eeq
as expected in \cite{NSVZ,CNZ}.
\section{Summary and conclusions}
We have systematically studied two-point functions built from light quark currents and gluon fields using the SVZ  $\oplus~1/q^2$ expansion and compared it with the predictions from some QCD holographic models, namely the Minimal Soft-Wall Model (MSW model), the Gauge/String Dual Model (GSD model) and the Hard-Wall Model (HW model). These two different approaches are expected to give a good parametrization of the non-perturbative aspect of QCD at moderate values of $q^2$ in the Euclidian region. We found that in most of the channels studied, the HW model shows a large departure 
from the SVZ idea as it does not have any power corrections. We also notice the equivalence between the MSW model and the GSD model analyzed by Ref. \cite{ANDREEV,ANDREEV1,ANDREEV2} from a constraint among the length parameters $c^2_i$ in the two models which is also obtained from their explicit relation with the tachyonic gluon mass. \\
The quadratic $1/q^2$-term introduced in \cite{CNZ,ZAK,SNZ}
in order to quantify the UV renormalon contributions and large order of the PT series appeared naturally in all examples studied within the MSW model and the GSD model with a sign and size consistent with a tachyonic guon obtained phenomenologically \cite{CNZ,SNI}. Pursuing the matching of the two approaches for higher dimension condensates, one notice that the  correlator of the $\bar qq$ scalar current is described consistently term by term by the two approaches until the $d=8$ condensates contribution reached in the SVZ-expansion. In the vector channel, the MSW model fails to reproduce the SVZ coefficient of the $d=6$ condensate contribution which vanishes in the MSW model due to some miraculous cancellation of its coefficients, while in the scalar gluonium channel, the MSW model cannot reproduce the sign of the Wilson coefficients of the $d\geq 4$ dimension condensates. In order to cure this
difficulty, we have briefly discussed an extension of the MSW model which is however less attractive than the MSW model due to its lesser predictivity.\\
From the relation of the dilaton parameter $c_i$ to the tachyonic gluon mass, we have also deduced a constraint among these parameters in different channels indicating that they are channel dependent but not universal for a given model. Then, we have deduced new values of
the hadronic mass ratios where the prediction for the scalar gluonium mass is in the range given by the substracted and unsubtracted gluonium sum rules \cite{SNG,VENEZIA,SNG0} and by lattice calculations with dynamical quarks \cite{HART}. We have also derived a hadronic hierarchy scale similar to the one obtained from QCD spectral sum rules analysis of different channels \cite{NSVZ,CNZ,SVZ,SNB}. 
\section*{Acknowledgement}
We thank Valya Zakharov for stimulating discussions.


\begin{thebibliography}{99}



\bibitem{SVZ} M.A. Shifman, A.I. and Vainshtein and V.I. Zakharov,
{\it Nucl. Phys.} {\bf B147} (1979) 385; M.A. Shifman, A.I. and Vainshtein and V.I. Zakharov,
{\it Nucl. Phys.} {\bf B147} (1979) 448.
\bibitem{SNB} For reviews, see
e.g., S. 
Narison, {\it QCD as a theory of hadrons,
Cambridge Monogr. Part. Phys. Nucl. Phys. Cosmol.} {\bf 17} (2002) 1
[hep-h/0205006]; ibid, {\it QCD
spectral sum rules ,  World Sci. Lect. Notes Phys.} {\bf 26}
(1989) 1;
ibid, { Acta Phys. Pol.} {\bf B26} (1995) 687; 
ibid, { Phys. Rept.} {\bf 84} (1982) 263; ibid, hep-ph/9510270 (1995).



\bibitem{CNZ} K. Chetyrkin, S. Narison and V.I. Zakharov, {\it Nucl. Phys.} 
{\bf B550} (1999)  353;
S. Narison and V.I. Zakharov, {\it  Phys. Lett.} {\bf B522} (2001) 266; 
\bibitem{ZAK} For reviews, see e.g.: V.I. Zakharov, {\it Nucl. Phys. Proc. Suppl.} 
{\bf 164} (2007) 240; S. Narison,  {\it Nucl. Phys. Proc. Suppl.} {\bf 164} 
 (2007) 225.
 
\bibitem{SNZ} S. Narison and V.I. Zakharov, {\it Phys. Lett.} {\bf B679} (2009) 355.

\bibitem{MALDA} J. M. Maldacena, {\it Adv. Theor. Math. Phys.} {\bf 2} (1998) 231.

\bibitem{WITTEN} E. Witten, {\it Adv. Theor. Math. Phys.} {\bf 2} (1998) 253; S. S. Gubser, I. R. Klebanov, A. M. Polyakov, {\it Phys.Lett.} {\bf B428} (1998) 105-114.

\bibitem{ANDREEV} J. Polchinski and M. Strassler, {\it Phys. Rev. Lett.} {\bf 88} (2002) 031601;
O. Andreev , {\it Phys. Rev.} {\bf D73} (2006) 046010.

\bibitem{ANDREEV1} O. Andreev , {\it Phys. Rev.} {\bf D73} (2006) 107901.

\bibitem{ANDREEV2} O. Andreev and V. I. Zakharov, {\it Phys. Rev.} {\bf D74} (2006) 025023;
{\it Phys. Rev.} {\bf D76} (2007) 047705.

\bibitem{ANDREEV3} O. Andreev, {\it Phys. Rev.} {\bf D82} (2010) 086012.

\bibitem{ERLICH} J. Erlich et al., {\it Phys.Rev.Lett.} {\bf 95} (2005) 261602.

\bibitem{JUGEAU3} A. Karch et al, {\it Phys. Rev.} {\bf D74} (2006) 015005


\bibitem{BROAD}D.J. Broadhurst and S.C. Generalis, {\it Phys. Lett.} {\bf B165} (1985) 175.



\bibitem{SNI}S. Narison, {\it Phys. Lett.} {\bf B300} (1993) 293; ibid {\bf B361} (1995) 121.

\bibitem{ZUO}
T. Huang and F. Zuo, {\it Chin. Phys. Lett.} {\bf 24} (2008) 3601.

\bibitem{JUG4}  
Karch et al., {\it JHEP} {\bf 1104} (2011) 066; F. Zuo, {\it Phys. Rev.} {\bf D82} (2012) 086011; S. J. Brodsky, G. F. de Teramond, H. G. Dosch,  [arXiv:1301.1651]. 

\bibitem{KIRKUN} A. Krikun, {\it Phys. Rev.} {\bf D77} (2008)126014. 

\bibitem{SNTAU}S. Narison, {\it Phys. Lett.} {\bf B673} (2009) 30.

\bibitem{LNT}G. Launer, S. Narison and R. Tarrach, {\it  Z. Phys.} {\bf C26}
(1984) 433.
\bibitem{fesr} R.A. Bertlmann, G. Launer and E. de Rafael, 
{ Nucl. Phys.} {\bf B250} (1985) 61; R.A. Bertlmann et al., 
{ Z.\ Phys.}  {\bf C39} (1988) 231.
\bibitem{YNDU}F.J. Yndurain, {\it Phys. Rept.} {\bf 320} (1999) 287-293 [hep-ph/9903457].
\bibitem{SNHeavy}S. Narison, {\it Phys. Lett.} {\bf B387} (1996) 162.
\bibitem{BELL}J.S. Bell and R.A. Bertlmann, {\it Nucl. Phys.} {\bf B227} (1983) 435;
R.A. Bertlmann, {\it Acta Phys. Austriaca} {\bf 53} (1981) 305;  R.A. Bertlmann 
and H. Neufeld, {\it Z. Phys.} {\bf C27} (1985)  437.
\bibitem{SNG2} S. Narison, {\it Phys. Lett.} {\bf B361} (1995) 121;
S. Narison,  {\it Phys. Lett.} {\bf B624} (2005) 223.
\bibitem{SNH10}S. Narison,  {\it Phys. Lett.} {\bf B693} (2010)  559; Erratum ibid 705 (2011) 544;
ibid, {\it Phys. Lett.} {\bf B706} (2011)  412; ibid, {\it Phys. Lett.} {\bf B707} (2012)  259. 

\bibitem{POMAROL}L. Da Rold and A. Pomarol, {\it Nucl. Phys.} {\bf B721} (2005) 79.

\bibitem{WEINBERG}S. Weinberg,  {\it Phys. Rev. Lett.} {\bf 18} (1967) 507; T. Das, V.S. Mathur and S. Okubo, {\it Phys. Rev. Lett.} {\bf 19} (1967) 470.

\bibitem{SUZUKI}K. Kawarabayashi and M. Suzuki, {\it Phys. Rev. Lett.} {\bf 16} (1966) 255;
Riazzudin and Fayazuddin, {\it Phys. Rev. } {\bf 147} (1966) 1071. 

\bibitem{FNR} E.G. Floratos, S. Narison and E. de Rafael, {\it Nucl. Phys.} {\bf B155} (1979) 115;
P. Pascual and E. de Rafael,  {\it Z. Phys.} {\bf C12} (1982)  127. 

\bibitem{SNWSR}S. Narison, {\it Z. Phys.} {\bf C14} (1982)  263; R.D. Peccei and J. Sol\`a,
{\it Nucl. Phys.} {\bf B281} (1987)1. 

\bibitem{KNECHT}M. Knecht and E. de Rafael, {\it Phys. Lett.} {\bf B424} (1998) 335.

\bibitem{JAMI2}Y. Chung et al.{\it Z. Phys.} {\bf C25} (1984)  151;  H.G. Dosch, 
Non-Perturbative Methods (Montpellier 1985);  
H.G. Dosch, M. Jamin and S. Narison, {\it Phys. Lett.} {\bf B220} (1989)  251.

\bibitem{COLAN}P. Colangelo et al., , {\it Phys. Rev.} {\bf D85} (2012) 035013; 
S. Nicotri, arXiv:1209.1929 (2012).
\bibitem{KNECHT2} M. Knecht et al., {\it JHEP} {\bf 0403} (2004) 035;
A. Czarnecki, W. J. Marciano and A. Vainshtein, {\it Phys. Rev.} {\bf D67} (2003) 073006

\bibitem{FJ} F. Jugeau [hep-ph/0902.3864]. 

\bibitem{HERY}H. Ratsimbarison, {\it Nucl. Phys. Proc. Suppl.} 
{\bf 207-208} (2010) 240.

\bibitem{JUGEAU}P. Colangelo et al, 
{\it Phys. Rev.} {\bf D78} (2008)055009.


\bibitem{NSVZ}V.A. Novikov et al.,
{\it Nucl. Phys.} {\bf B191} (1981) 301. 

\bibitem{SNG}S. Narison, {\it Phys. Rev.} {\bf D 73} (2006) 114024;
S. Narison,  {\it Nucl. Phys. } {\bf B509} (1998) 312; Nucl.Phys.Proc.Suppl. 64 (1998) 210-219. 

\bibitem{JUGEAU2}P. Colangelo et al.,
{\it Int. J. Mod. Phys.} {\bf A24, 22} (2009) 4177; P. Colangelo et al., {\it Phys.Lett.} {\bf B652} (2007) 73-78.

\bibitem{FORKEL1}H. Forkel, {\it Phys. Rev.} {\bf D78} (2005) 054008.

 \bibitem{VENEZIA} S. Narison and G. Veneziano, {\it Int. J. Mod. Phys.} {\bf A4} (1989) 2751. 
 
 \bibitem{SNG0}S. Narison, {\it Z. Phys.} {\bf C26} (1984) 209; E. Bagan and T.Steele, {\it Phys. Lett.} {\bf B243} (1990) 413.
 
  \bibitem{MINK}P. Minkowski, W. Ochs, {\it  Eur. Phys. J.} {\bf  C9} (1999) 283; 
 A. Bramon and S. Narison, {\it Mod. Phys. Lett.} {\bf A4} (1989) 1113.

  \bibitem{MENES} G. Mennessier, S. Narison, W. Ochs, {\it Phys. Lett.} {\bf B665} (2008) 205; 
{\it Nucl. Phys. Proc. Suppl.} {\bf 238} (2008) 181; 
G. Mennessier, P. Minkowski, S. Narison, W. Ochs, HEPMAD 07 Conference,  SLAC Econf  C0709107, {\it arXiv: 0707.4511 [hep-ph] } (2007); R. Kaminski, G. Mennessier and S. Narison, {\it Phys. Lett.} {\bf B680} (2009) 148; G. Mennessier, S. Narison and X.-G. Wang, {\it Phys. Lett.} {\bf B688} (2010) 59;  ibid, {\it Phys. Lett.} {\bf B696} (2011) 40.
 
 \bibitem{HART} A. Hart et al. [UKQCD Collaboration], {\it Phys. Rev.} {\bf D74} (2006) 114504.

 \bibitem{LATT}G. S. Bali et al. [UKQCD Collaboration], {\it Phys. Lett.} {\bf B309} (1993) 378; 
C. J. Morningstar and M. J. Peardon, {\it Phys. Rev.} {\bf D 60} (1999) 034509; Y. Chen et al., {\it Phys. Rev.} {\bf D}73 (2006) 014516.

\bibitem{OCHS}W. Ochs, arXiv:1301.5183 (2013).

\bibitem{FORKEL2}C. Csaki, H. Ooguri, Y. Oz and J. Terning, {\it JHEP} {\bf 9901} (1999) 017; 
R. de Mello Koch, A. Jevicki, M. Mihailescu and J. P. Nunes, {\it Phys. Rev.} {\bf D58} (1998) 105009;
M. Zyskin, {\it Phys. Lett.} {\bf B439} (1998) 373; 
J. A. Minahan, {\it JHEP} {\bf 9901} (1999) 020;
R. C. Brower, S. D. Mathur and C.-I Tan, {\it Nucl.Phys.} {\bf B587} (2000) 249;
R. Apreda, D. E. Crooks, N. J. Evans and M. Petrini, {\it JHEP} {\bf 0405} (2004) 065; 
U. Gursoy, E. Kiritsis, and  F. Nitti, {\bf JHEP} {\bf 0802} (2008) 019;
H. Forkel, {\it Phys. Rev.} {\bf D78} (2008) 025001.

\bibitem{KATAEV}A.I. Kataev, N.V. Krasnikov and A.A. Pivovarov, {\it Nucl.Phys.} {\bf B198} (1982) 508; Erratum-ibid, {\bf B490} (1997) 505.

\end{thebibliography}
\end{document}